# The Origin of Organic Magnetoresistance Using Time-Domain Magnetic Spectroscopy

Paul Ben Ishai[1*], Roshlin Kiruba[1] and Amos Bardea[2]

[1]Department of Physics, Ariel University, Ariel 4700000, Israel

[2]Department of Electrical and Electronics Engineering, Holon Institute of Technology, Holon 5810201, Israel, Israel

*corresponding author:  Paulbi@ariel.ac.il

## Abstract

Organic Magnetoresistance is defined as the change of resistance in an organic material, such as a conducting polymer, as a function of an imposed magnetic field.  We demonstrate this effect in a Polypyrrole/ Polydimethylsiloxane complex by using a novel magnetic pulse system.  The frequency spectrum of the current flowing through the sample reveals equally spaced reversed peaks in the current.  We show that these peaks happen at the Lamour frequency for the dominant charge carrier of the system, namely polarons.  This posits the origin Organic magnetoresistance as simple Rabi Oscillations rather than mechanisms based of bipolaron formation and singlet-triplet conversions. We directly estimate the effective polaronic mass in this complex.  A semi classical theoretical approach is suggested to explain this effect as direct spin flipping in a time transient magnetic field.  This is the first time that such an experimental approach has been applied in this field and the first time the effective polaronic mass has been measured in a conducting polymer.

## Magnetic Resistance in Conducting Polymers

Organic Magnetoresistance (OMAR), the change of resistance in an organic material as a function of an imposed magnetic field, has been noted in conjugated conducting polymers for more than a decade  [1]. As opposed to magnetoresistance in inorganics OMAR is characterized by independence from the magnetic field direction, occurs at room temperature rather than at Helium temperatures and at relatively modest field strengths  [2]. For conjugated polymers, the conduction mechanism has been traced to delocalized π-electrons interacting with the underlying polymer backbone to form polarons  [3,4], but the microscopic origin of OMAR is still under debate, as there can be a number of different mechanisms involved [5,6].   Most researchers agree that the hyperfine interaction between delocalized $\pi$ electrons and the hydrogen ion responsible for the $\pi$-bond allows the formation of polarons, bi-polarons, excitons [1,7] and some form of spin correlation between charge carriers [8]. The action of the external field is to break this interaction and effectively align spin moments. This in turn brings the fermionic nature of the charge carrier comes into play, manipulating the impedance of the sample. The nature of this interaction is still under debate with a number of possibilities put forward: electron–hole pair formation [9], triplet–polaron interaction [10], singlet–triplet interconversion [11], triplet–triplet annihilation [12] and Pauli spin blockade [13] to name a few. These mechanisms require the existence of two charge carriers at the same molecular site and a lattice distortion to lower the repulsive Coulombic interaction between them [8,13,14]. The hyperfine interaction of their spins with the local nuclear magnetic field then allows bipolaron formation. We show below, however, that invoking bipolaron formation may not be necessary to explain OMAR in conjugated polymers. It maybe enough to invoke simple Rabi-like oscillations. Typically, the gauge of OMAR is defined by

$$\frac{\Delta Z}{Z(0)} = \frac{Z(B)-Z(0)}{Z(0)} \qquad (1)$$

Where $Z(B)$ is the measured impedance under the influence of a static external magnetic field and $Z(0)$ is the field free impedance. In this definition OMAR can be positive or negative, depending on the action of the magnetic field to facilitate or impede charge carriers. Organic Magnetoresistance (OMAR) was first observed in materials using a spin valve architecture, derived from inorganic spintronics [1]. In this configuration, spin–polarized carriers were injected into the organic spacer sandwiched between ferromagnetic electrodes [8]. The interest in OMAR has surged with the replacement of ferromagnetic electrodes by non-ferromagnetic metals conventionally used in organic electronics. This approach has achieved significant magnetoresistance (>10%) even in low magnetic fields ($B \sim 10\ mT$) at room temperature [5] in a ITO/PEDOT/polyfluorene/Ca vertical diode. Immense OMAR (>1000%) has been measured in devices based on organic blends exhibiting thermally activated delayed fluorescence [15]. Such a magnetoresistance effect is amongst the largest for any non-magnetic bulk material. Broadly speaking the relevant conducting polymers fall into 4 categories (illustrated in Figure 1); (1) intrinsically conducting such as Polypyrrole (PPy), (2) Dopants/polymer charge transfer complexes like Sodium Hydride in Polyaniline to n-dope the polymer, (3) Organomettalic polymers like Polyvinylferrocene [16], and (4) conducting polymer composites such as carbon nanotubes in polyethylene [17]. OMAR has been demonstrated in Polypyrrole [1,18]. The polymer is formed by the oxidative polymerization of pyrrole, a heterocyclic aromatic compound with the formula $C_4H_4NH$ arranged in a 5 membered ring [19] and it is notable for have a distinct energy gap of 1.88 eV [20]. We investigate OMAR in PPy suspended in an elastomer matrix by mixing PPy in Polydimethylsiloxane (PDMS) [21]. This allows the formation of flexible material that maintains its conductivity without succumbing to the known brittleness of pure PPy [22]. We employ a novel time domain method to investigate OMAR effects.

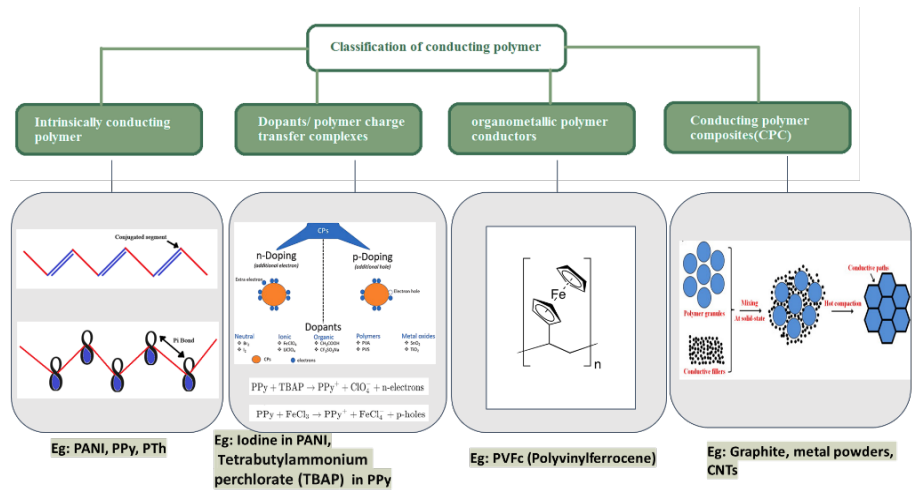

Figure 1 - The four categories of conducting polymers, based on images from Ref *[3]*, *[4]* (under common license CC BY-NC-ND 4.0) and *[23]*

Time domain approaches to spectroscopies, specifically dielectric spectroscopies, were quite prevalent in the 1970s. Pioneers like R. Cole [24] and Yu. Feldman [25] produced spectrometers based on TDR methods, a method first promulgated (in the West at least) in a Technical paper produced by Hewlett Packard [24,26]. These techniques are based on linear response theory, whereby the shape and form of a reflection of an impulse signal from a mismatch in impedances will inherently contain the characteristics of that mismatch, in short the difference in permittivities on each side of the mismatch [25,27,28]. The method is readily applicable to magnetic pulses rather than electric. There are distinct advantages in using the time domain; the measurement is completely in the real plane and, depending on the sampling rate and window, the frequency response over a broad band of frequencies is gathered simultaneously. However, the resultant signal is a convolution of the response of the material to the impulse and direct deconvolution is not a simple task. This can be overcome by the direct Fourier image of the input and output signals whereby the convolution becomes a simple multiplication of signals in the frequency domain. We apply the same rational to the investigation of Magnetoresistance. In general, the current density, $j(t)$, in polymer is a response, $\alpha(t)$, to an impinging time based magnetic field, $B(t)$ and would be formally [29,30] described by

$$j(t) = \int_{-\infty}^{t} B(t')\alpha(t-t')dt' \qquad (2)$$

Such an approach does not directly allow for the vectorial nature of the interaction and this must be arranged by careful experimentation. Once done it is advantageous to measure in the time domain but analyse the work in the frequency domain. We show below that the Fourier images of the current and impedance response reveal a characteristic quantised spin/field interaction at room temperature. The results are explained by semi-classical model that accounts for the vectorial nature of the interaction. The simplicity of the approach allows one to measure the signal using only an oscilloscope.

## Methods and Materials

Polydimethylsiloxane (PDMS) in the form of Sylgard 184 was purchased from Dow Chemicals. Polypyrrole was purchased from Sigma in powder form and mixed to PMDS at a 3% weight ratio. Copper strips as electrodes were placed at the bottom of a square mold, approximately 1 cm square and the polymer mix cast into it. The samples were degassed for 30 minutes and allowed to cure for 24 hours. The magnetoresistance of such films, their sensitivity to mechanical deformation and temperature have been classified previously [21]. Samples were also made with additional Nickel nanoparticles (NiNP), 500 nm diameter from Sigma Aldritch, added in low (5% weight ratio) or high (30% weight ratio) concentrations. Silica (500 nm diameter, Sigma Aldritch) and Iron oxide in the alpha phase (500 nm diameter, Sigma Aldritch) were used as well at 10% weight ratio. Cross sectional EDX spectroscopy shows that the NiNP are homogeneously distributed throughout the sample [21] and we assume a similar distribution for the other nanoparticles.

At room temperature NiNP are ferromagnetic [31] and should contribute to local magnetic fields in the sample. Silica particles were added to probe possible space and local field effects and iron oxide particles (hematite α-$Fe_2O_3$) as n-dopants.

The experimental system consists of a sampling oscilloscope (Keysight DSOX 1204G 200 MHz), a function generator acting as a trigger (Rigol DG1022), a current pulser produced in house and capable of producing a repeating current pulse of up to 100 A with a duration of approximately 1 millisecond and a copper solenoid with 9 turns. The current pulser can be considered as a glorified RC circuit and its pulse can be modelled by $I(t) =$

$S(t - t_0)I_0 \exp[-(t - t_0)/\tau_I]$, where $S(t - t_0)$ is a heavyside function, $I_0$ is the current amplitude and $\tau_I$ is the characteristic decay time of the pulse. The system is schematically represented in Figure 2. To measure the impedance the sample is set in series with a passive resistor. A constant voltage over the circuit is maintained by a 3 volt battery. The sample is held at the opening of the solenoid and the magnetization is measured by a hall sensor (Honeywell SS495A1 through hall sensor) with a maximum sampling rate of 20 kHz and a sensing area of 50 $\mu m^2$. The time varying current through the sample is measured as the voltage variation over the resistor. Under these conditions the impedance of the sample is given by

$$Z(\omega) = \frac{\mathcal{L}[V_S]}{\mathcal{L}[I]} = R\left[\frac{3}{i\omega\mathcal{L}[V_R(t)]} - 1\right] \tag{3}$$

Where $\mathcal{L}[\ ]$ is the Fourier operator, $V_R(t)$ is the measured resistor voltage and $R$ is the passive resistance. The relative permeability of the sample during the magnetic pulse is calculated by the ratio of signals from the hall sensor for the sample and for the empty coil, $\mu(\omega) = H(\omega)/B(\omega)$. To account for non-linearity of the hall sensor its voltage response is first calibrated against the calculated magnetic field of the coil using the measured coil current, $I(t)$, and the solenoid equation, $B(t) = \mu_0 nI(t)$, where $\mu_0$ is the permeability of free space and $n$ is the coil density.

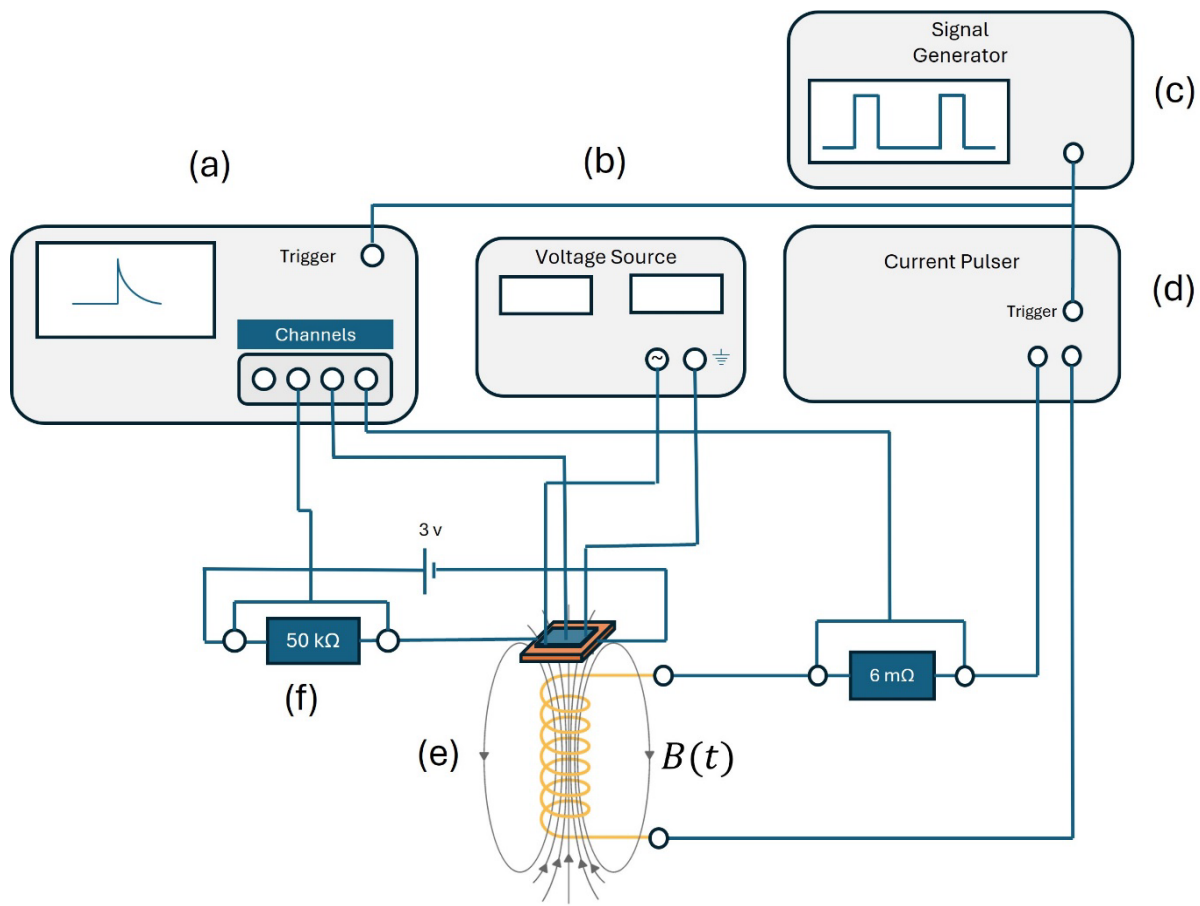

*Figure 2 - A schematic representation of the measurement setup; (a) A sampling oscilloscope, (b) A voltage source powering a hall sensor, (c) A function generator to trigger (d) a current pulser producing pulses of 1 millisecond duration and an amplitude of 100 A, (e) a copper solenoid with 10 turns and (f) a voltage divider circuit consisting of a 3v battery , a passive resistor and the sample. The sample is physically placed at the opening of the coil and has a hall sensor attached to its upper side. The variations of current passing through the sample, as the transient magnetic field affects the sample impedance, are registered directly by the oscilloscope along with the current pulse through the solenoid and the hall sensor voltage.*

The relative permeability of the samples was measured at frequencies lower than 100 Hz in order to avoid the intrinsic decay of the current pulser. The results show that the basic PPy/PDMS sample is diamagnetic with a static relative permeability of $\mu = 0.94 \pm 0.01$. The addition of NiNP (low density sample) led to a reduction of the relative permeability to $\mu = 0.82 \pm 0.01$.

The signals were averaged over 30 pulses, Fourier transformed and are presented in Figure 3. Figure 3(a) shows the voltage drop as a function of time across the passive resistor of the circuit described by Figure 2, as measured by the sampling oscilloscope during the magnetic pulse experienced by the sample.

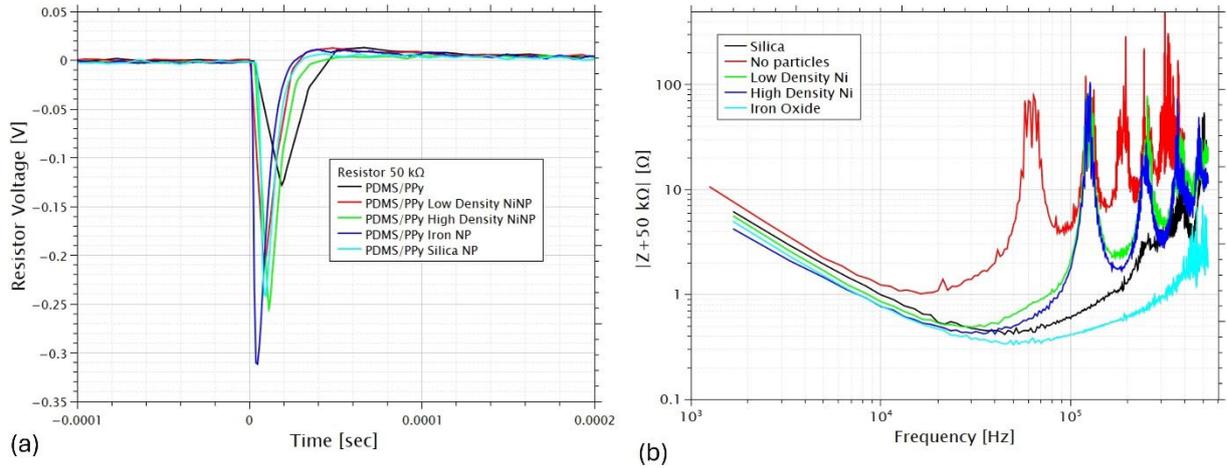

*Figure 3 - (a) the measured voltage drop across the passive resistor (50 Ω) for 5 different conducting polymer systems, variously doped with nanoparticles and (b) the impedance through the samples as calculated by equation (4).*

Figure 3(b) presents the impedance of the samples as calculated by equation (3). The lower frequencies are characterized by a power law slope, while the higher frequencies show equally spaced peaks on a rising background. We note that the main contribution of nanoparticles to the polymers is a reduction in the overall impedance, indication the addition of charge carriers to the system. This holds true even for the addition of silica particles. The spacing of the peaks does depend on the value of the passive resistor, as can be seen in Figure 4. The characteristic frequency with a 50 kΩ resistor is $f_0$=0.66 ±0.09 MHz for a sample without nanoparticle doping and 5.00±0.03 kHz when using a 50 Ω resistor. The values of the characteristic frequencies (50 kΩ) for the doped samples are $f_0$=0.125±0.04 MHz (Silica NP), 0.129±0.001 MHz (NiNP at low density) and 0.127±0.003 MHz (NiNP at high density).

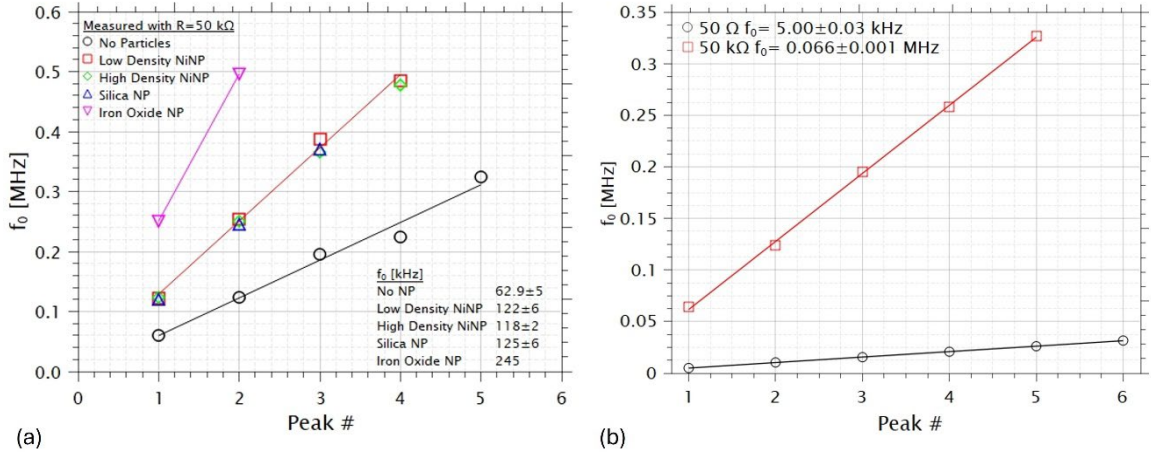

*Figure 4 - (a) The peak positions as taken from Figure 3(b). The underlying characteristic frequency, $f_0$ for the basic conducting polymer is 0.066 MHz. Once doped with a nanoparticle this frequency is approximately doubled. (b) The value of the characteristic frequency is directly dependent on the passive resistor value. The data is measured for the pure PPy sample with a 50Ω (red squares) and a 50 kΩ (black circles) resistor. The change is traced to a modification of the effective polaron mass caused by differing bias voltages across the sample. Linear regression is used to fit the Lamour frequency for each sample.*

As described above the mechanism of charge transport in PPy is the hopping of delocalized π-electrons along the polymer backbone, forming polarons [32].

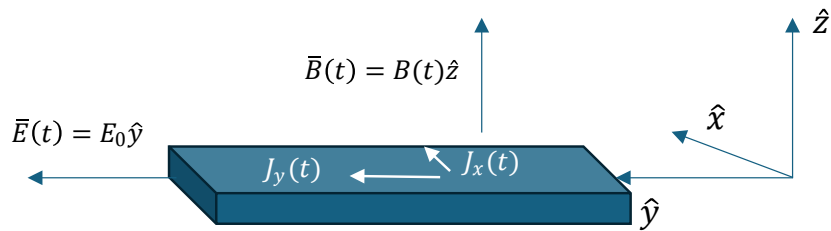

*Figure 5 - Schematic of the frames of reference for the measurement setup.*

The effect of a magnetic field on the charge carriers of the sample can be modelled by considering 3 forces at work on the charge carriers, an approach reminiscent of the treatment of the Integral Quantum Hall Effect [33]. If one considers the sample as a slab in the x-y plane with a current passed through in the x direction and a time dependent magnetic field along the z axis, then the forces acting on the charge carriers in the sample are: $\boldsymbol{F_1} = q\boldsymbol{E}$, the Electrical force, $\boldsymbol{F_2} = q(\boldsymbol{v} \times \boldsymbol{B}) = \frac{1}{n_e}(\boldsymbol{J} \times \boldsymbol{B})$, the Lorentz force, where $n_e$ is the number density of charge carriers and $\boldsymbol{J}$ is the current density, and $\boldsymbol{F}_3 = \nabla(\boldsymbol{m} \cdot \boldsymbol{B})$, the gradient of the interaction energy of the magnetic field with the charge carrier's magnetic moment. Considering these three forces as effective electric fields, the current density can be defined by $\boldsymbol{J} = \sigma \boldsymbol{E} = \frac{\sigma n_e}{q}(\boldsymbol{F}_1 + \boldsymbol{F}_2 + \boldsymbol{F}_3)$, where $\sigma$ is the specific conductivity tensor. Explicitly this can be written as

$$\boldsymbol{J} = \sigma \left( n_e \boldsymbol{E} + \frac{1}{q}(\boldsymbol{J} \times \boldsymbol{B}) + \frac{n_e}{q} \nabla(\boldsymbol{m} \cdot \boldsymbol{B}) \right) \tag{5}$$

The first two terms can be considered classically but the third term is a pure quantum mechanical element. Taking into consideration the axis definitions, illustrated in Figure 5, and assuming that the electric field in the y direction is dominant, the components can be written as $\overline{\boldsymbol{B}}(\omega) = B(\omega)\hat{z}$, $\boldsymbol{J} = J_x \hat{x} + J_y \hat{y}$ and $\boldsymbol{E} = E_x \hat{x} + E_y \hat{y} \approx E_y \hat{y}$ and the final expression can be simplified to

$$J_y = n_e \sigma E_y - \frac{\sigma^2}{q^2} J_y B_z^2(\omega) - i\omega \sigma n_e^2 \frac{1}{J_y} \langle m_z B_z \rangle \tag{6}$$

Where use has been made of the identity $\left( \frac{\partial}{\partial x}, \frac{\partial}{\partial y}, \frac{\partial}{\partial z} \right) \rightarrow \frac{\partial}{\partial t}\left( \frac{1}{v_x}, \frac{1}{v_y}, \frac{1}{v_z} \right) \rightarrow i\omega q \left( \frac{n_e^x}{J_x}, \frac{n_e^y}{J_y}, \frac{n_e^z}{J_z} \right)$, the velocities in this case are the drift velocities of the charge carriers and can be related to the current densities by the relationship $J_{x,y,z} = q n_e v_{x,y,z}$, where $q$ is the elemental charge and $n_e$ is the charge density. The final step brings the expression into the frequency domain by Fourier transform. The third term in equation (6) is the expectation value of the energy of the spin- B field interaction. However, a direct calculation of the energy spectrum of a spin system in decaying magnetic field with a fixed orientation would result in a featureless energy spectra as spin states would remain static, gaining only phase during the time pulse of the magnetic field. It would take the small, but necessary. perturbation of the spin particle generated by the time decaying electric field in the x direction to induce flipping between states and spectra noted in Figure 3(b). Heuristically, this can be appreciated as follows. The source of this perturbation is of course the classic Lorentz force generated by the magnetic pulse because of the motion of the charge carriers. This motion leads to a time varying electric $E_x(t) \sim \exp(-t/\tau)\hat{x}$ that would generate a time varying current and therefore a complimentary magnetic field, $B_y(t) \sim \exp(-t/\tau)\hat{y}$. This provides a perturbation to the dominant spin states. One can account for this by modifying equation 6 to read

$$J_y = n_e \sigma E_y - \frac{\sigma^2}{q^2} J_y B_z^2(\omega) - i\omega \sigma n_e^2 \frac{1}{J_y} \langle \boldsymbol{m} \cdot (B_y(t)\hat{y} + B_z(t)\hat{z}) \rangle \tag{7}$$

As both components of the effective magnetic field decay exponentially with the same time constant the Hamiltonian can be written as $H(t) = \left(\frac{\hbar\omega_0}{2}\sigma_z + \frac{\hbar}{2}\xi\sigma_y\right)\exp(-t/\tau)$, where $\omega_0 = \frac{qB_0}{2m_{eff}}$ is the Lamor frequency and $\xi \ll \omega_0$. This Hamiltonian can be solved exactly using Pauli matrices [33,34]. The time instantaneous energy of such a system can be written as

$$E_\pm(t) = \pm\hbar\omega(t) = \pm\frac{\hbar}{2}\omega_0 \exp(-t/\tau)\left(1 + \frac{\xi^2}{2\omega_0^2}\right) \quad (8)$$

Equation 8 suggests an energy spectrum for flipping between spin energy states, $\Delta E(\omega) = |E_+ - E_-|$, is then

$$\Delta E(\omega) = \hbar|\mathcal{L}[\Delta\omega(t)]| \approx \hbar\frac{\omega_0}{1+(\tau\omega)^2} \quad (9)$$

This would be the classic Lamor frequency, akin to a Rabi-like oscillation [35], modulated by a Lorentzian with a FWHM dictated by the decay time of the magnetic pulse. For a spin system this would give periodic changes in the impedance by $\langle m \cdot B \rangle \sim \langle m_z B_z \rangle = \hbar\omega_0/(1+\tau^2(\omega - N\omega_0)^2)$, where $N$ is a positive integer, the Larmour frequency is given by $\omega_0 = \frac{qB_0}{2m_{eff}}$, where $m_{eff}$ is the effective charge carrier mass. The solution to equation 6 is given by

$$J_y = \frac{n_e\sigma E + \sqrt{(n_e\sigma E)^2 \pm i4\omega\sigma n_e^2\hbar\omega_0/(1+\tau^2(\omega-N\omega_0)^2)\left(1+\frac{\sigma^2}{q^2}B^2\right)}}{2\left(1+\frac{\sigma^2}{q^2}B^2\right)} \quad (10)$$

Equation (10) can be linearized to

$$J_y = \frac{n_e\sigma E}{\left(1+\frac{\sigma^2}{q^2}B^2\right)} \pm i\frac{n_e\omega}{E}\hbar\omega_0/(1+\tau^2(\omega-N\omega_0)^2) \quad (11)$$

The absolute value of the current density is then

$$|J_y|^2 = \left[\frac{n_e\sigma E}{\left(1+\frac{\sigma^2}{q^2}B^2\right)}\right]^2 - \left[\frac{n_e\omega}{E}\hbar\omega_0\right]^2 \frac{1}{1+\tau^2(\omega-N\omega_0)^2} \quad (12)$$

Inverting equation (12) brings one to an expression for the impedance $Z = \left(\frac{l}{A}\right)\frac{E}{J}$, of the sample (l is the sample length and A its cross-sectional area). However, it is easier to analyse the current, plotted in Figure 5, and this follows the general form of equation (12) with periodic drops in the absolute current values passing through the sample, as spins switch between parallel (high energy) or antiparallel (low energy) alignment to the B field direction.

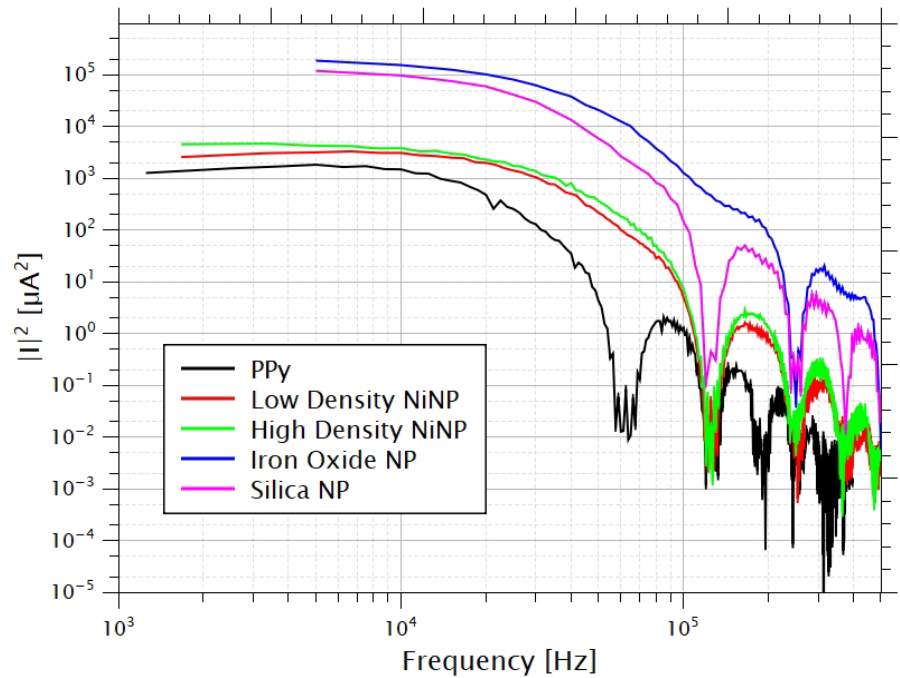

Figure 6 - The current through the samples. The signal shape closely follows the predicted behaviour of equation 12, with the magnetic field of the form $B(\omega) = B_0/(1 + i\omega\tau_B)$, where $B_0$ is the field amplitude (0.157 T for NiNP and pure PPy, 0.183 T for Iron and Silica NP) and $\tau_B$ is the characteristic decay rate of the current pulse. The predicted characteristic frequencies corresponding to $\omega_0$ of equation 12 are the same as those noted in the impedance spectra. Peak widths correspond to the expected time constant of the B field decay, $\tau \approx 1\ ms$.

The peak positions in Figure 6 naturally correspond to those of Figure 4(a) and correspond to the integer multiples of the Lamour frequencies. This interpretation of OMAR in polypyrrole requires no bipolaron formation or exotic singlet – triplet transitions. The origin of OMAR is simply Rabi Oscillations.

In Figure 4(b) one can note that using a smaller resistance in the circuit has a direct effect on the Lamour frequency $\omega_0 = 2\pi f_0 = \frac{qB_0}{2m_{eff}}$,. For the PPy sample with R=50 kΩ (Figure 4(a)) the Lamour frequency was $f_0 = 6.29 \pm 0.5\ kHz$, whereas for R=50 Ω one has $f_0 = 5.00 \pm 0.03\ kHz$. For the doped samples the Lamour frequency is very similar; $f_0 = 0.167 \pm 0.003\ MHz$ for High Density NiNP, $f_0 = 0.129 \pm 0.001\ MHz$ for Low

Density NiNP and $f_0 = 0.125 \pm 0.004\ MHz$ for Silica NP. However, Iron Oxide NP gave $245\ MHz$. The influence of the resistor value on the Lamour frequency can be traced to the effective mass of the charge carriers, $m_{eff}$. The particle's effective mass [36] is defined in terms of the band curvature in $k$ space. Further coupling between the charge carrier and the underlying lattice leads to a modification and this is accounted for by a coupling parameter $\alpha$,

$$m^* = \frac{1}{\hbar^2}\frac{\partial^2 E}{\partial k^2}\ , \qquad \frac{m^{**}}{m^*} = 1 + \alpha + \alpha^2 \qquad (1)$$

Estimates of the value of the effective mass vary. Initial theoretical considerations by Landau [37] posited effective masses of up to 430 $m_e$ in rigid lattices like NaCl, due to strong coupling. Later theoretical approaches, like the Fröhlich (large polarons) [38] and Holstein (small polarons) [39,40] models coupled with Feynman's path integral approach [41], led to effective masses of up to 1000 $m_e$ [42] (with $\alpha \sim 12$). In the simple circuit used here to measure the current (see Figure 2) the resistor acts as a voltage divider. Initially the sample is in steady state with a constant current flowing through it. As only the change in the current due to the impinging magnetic field is relevant to the main discussion, Figure 3(a) shows the steady state condition offset as a zero voltage. As PPy is a direct gap semiconductor with $E_g = 1.88\ eV$ [20,43], the effective mass of charge carriers, $m_{eff}^{-1} \propto \partial^2 E/\partial k^2$, promoted to the LUMO band (conduction band) would depend on the curvature of the LUMO and HOMO bands at the Γ point in the first Brillion zone of the E-k diagram (see Figure 12 of reference [20]). One could assume that a bias voltage across the sample, which is proportional to the sampling resistor, would perforce lead to band bending at the electrode/polymer interface and a consequent change in the curvature, hence a different effective mass. However, it could also be that the coupling coefficient, $\alpha$, which is influenced by the conduction process (polaron hoping along the polymer backbone, tunneling or hopping Ppy intersections and other possible scenarios) is more likely directly affected by the differing bias voltages. The same situation would arise when NPs are used to enhance the conductivity, creating a composite. The ratio in effective mass between the undoped PPy sample and those doped by NPs, including the inert Silica, follows the respective ratios of the Lamour frequencies.

$$\frac{m_{eff}^{doped}}{m_{eff}^{PPy}} = \frac{\omega_0^{PPy}}{\omega_0^{doped}} = \frac{0.066}{0.127} \approx \frac{1}{2} \qquad (10)$$

This points to the local field of the nanoparticle as the modifying factor in the band structure. For instance, the Silica nanoparticle Zeta potential is negative and about -20 mV [44], implying significant local fields, even if its contribution to charge carriers is minimal. Notedly the ferromagnetic nature of Nickel nanoparticles has little effect on the magnetic behavior of the samples.

The amplitude of the magnetic field, $B_0$, is measured as a time dependent voltage drop across the 6 mΩ resistor in Figure 2. From fitting its value was $B_0 = 0.158\,T$ for the measurements of plain PPy, high and low density NiNP, and $B_0 = 0.183\,T$ for the Silica and the Iron Oxide NP samples Using these values we can derive a direct estimation of the effective mass of the charge carrier, polaron in this case, to be

$$m_{eff} = \frac{qB_0}{\omega_0} \quad (10)$$

Taking the values of $\omega_0$ when using a 50 kΩ resistor from Figure 4(a), one has for undoped PPy a polaron effective mass of $m_{eff} = (70 \pm 6) \times 10^3\,m_e$ and for doped PPy (Low density NiNP) $m_{eff} = (36 \pm 2) \times 10^3 m_e$. For Silica nanoparticles one has $m_{eff} = (41 \pm 2) \times 10^3 m_e$ and $m_{eff} = 21 \times 10^3 m_e$ for Iron oxide NP. Using a 50 Ω resistor the effective mass of a polaron in undoped PPy is $m_{eff} = (1030 \pm 6) \times 10^3 m_e$ These results tally with estimates for effective polaron masses calculated for such organic semiconductors using Langevin molecular dynamics [45]. However, our results can be considered as a direct measurement.

## Conclusions

The quantum aspects of magnetoresistance in conducting polymers are readily accessible at room temperature when a time dependent magnetic field is used. To the best of our knowledge this is the first time the relatively old technique of time domain response has been applied to magnetic spectroscopy. We have revealed a pure spin field interaction of the main charge carrier, delocalized π-electrons forming polarons, along the polymer backbone of Polypyrrole. The impedance spectra reveal periodic peaks corresponding to energy spin flips in the spin – B field interaction, as spins jump from parallel to anti-parallel orientations. We posit this as the origin of magnetoresistance, rather than spin blockades or mechanisms based on bipolaron formation. Furthermore, we have presented a theoretical, semi-classical model that explains the frequency representation of the current density. One of the main achievements of the model is the dependence of the Lamour frequency on the effective bias field across the sample, dictated by the in series resistor.  As this is reciprocal to the effective polaron mass, it is a simple window to its direct measurement in conducting polymer systems. The values we have obtained for the effective polaron mass confirm estimations derived from simulation studies.